\documentclass[sigconf]{acmart}

\AtBeginDocument{%
  \providecommand\BibTeX{{%
    \normalfont B\kern-0.5em{\scshape i\kern-0.25em b}\kern-0.8em\TeX}}}

\copyrightyear{2024}
\acmYear{2024}
\setcopyright{acmlicensed}
\acmConference[UIST '24]{The 37th Annual ACM Symposium on User Interface Software and Technology}{October 13--16, 2024}{Pittsburgh, PA, USA}
\acmBooktitle{The 37th Annual ACM Symposium on User Interface Software and Technology (UIST '24), October 13--16, 2024, Pittsburgh, PA, USA} \acmDOI{10.1145/3654777.3676336}
\acmISBN{979-8-4007-0628-8/24/10}

\usepackage{enumitem}
\usepackage{multirow}
\usepackage{color}
\usepackage{glossaries}

\newacronym{mr}{MR}{Memory Reviver}

\begin{document}

\title[Memory Reviver]{Memory Reviver: Supporting Photo-Collection Reminiscence for People with Visual Impairment via a Proactive Chatbot}

\author{Shuchang Xu}
\affiliation{
\institution{Hong Kong University of Science and Technology}
\city{Hong Kong}
\country{China}
}
\orcid{0000-0002-7642-9044}
\email{xschci@gmail.com}

\author{Chang Chen}
\affiliation{
\institution{Hong Kong University of Science and Technology}
\city{Hong Kong}
\country{China}
}
\orcid{0009-0008-5876-7219}
\email{cchenda@connect.ust.hk}

\author{Zichen Liu}
\affiliation{
\institution{Hong Kong University of Science and Technology}
\city{Hong Kong}
\country{China}
}
\orcid{0009-0009-0078-6238}
\email{zliucz@connect.ust.hk}

\author{Xiaofu Jin}
\affiliation{
\institution{Hong Kong University of Science and Technology}
\city{Hong Kong}
\country{China}
}
\orcid{0000-0002-7239-3769}
\email{xjinao@connect.ust.hk}

\author{Lin-Ping Yuan}
\authornote{Corresponding author.}
\affiliation{
\institution{Hong Kong University of Science and Technology}
\city{Hong Kong}
\country{China}
}
\orcid{0000-0001-6268-1583}
\email{lyuanaa@cse.ust.hk}

\author{Yukang Yan}
\affiliation{
\institution{University of Rochester}
\city{New York}
\country{United States}
}
\orcid{0000-0001-7515-3755}
\email{yukang.yan@rochester.edu}

\author{Huamin Qu}
\affiliation{
\institution{Hong Kong University of Science and Technology}
\city{Hong Kong}
\country{China}
}
\orcid{0000-0002-3344-9694}
\email{huamin@cse.ust.hk}

\renewcommand{\shortauthors}{Xu and Chen, et al.}

\begin{abstract}
Reminiscing with photo collections offers significant psychological benefits but poses challenges for people with visual impairment (PVI). Their current reliance on sighted help restricts the flexibility of this activity. In response, we explored using a chatbot in a preliminary study. We identified two primary challenges that hinder effective reminiscence with a chatbot: the scattering of information and a lack of proactive guidance. To address these limitations, we present Memory Reviver, a proactive chatbot that helps PVI reminisce with a photo collection through natural language communication. 
Memory Reviver incorporates two novel features: (1) a \emph{Memory Tree}, which uses a hierarchical structure to organize the information in a photo collection; and (2) a \emph{Proactive Strategy}, which actively delivers information to users at proper conversation rounds. Evaluation with twelve PVI demonstrated that Memory Reviver effectively facilitated engaging reminiscence, enhanced understanding of photo collections, and delivered natural conversational experiences. Based on our findings, we distill implications for supporting photo reminiscence and designing chatbots for PVI.

\end{abstract}




\begin{teaserfigure}
  \includegraphics[width=\textwidth]{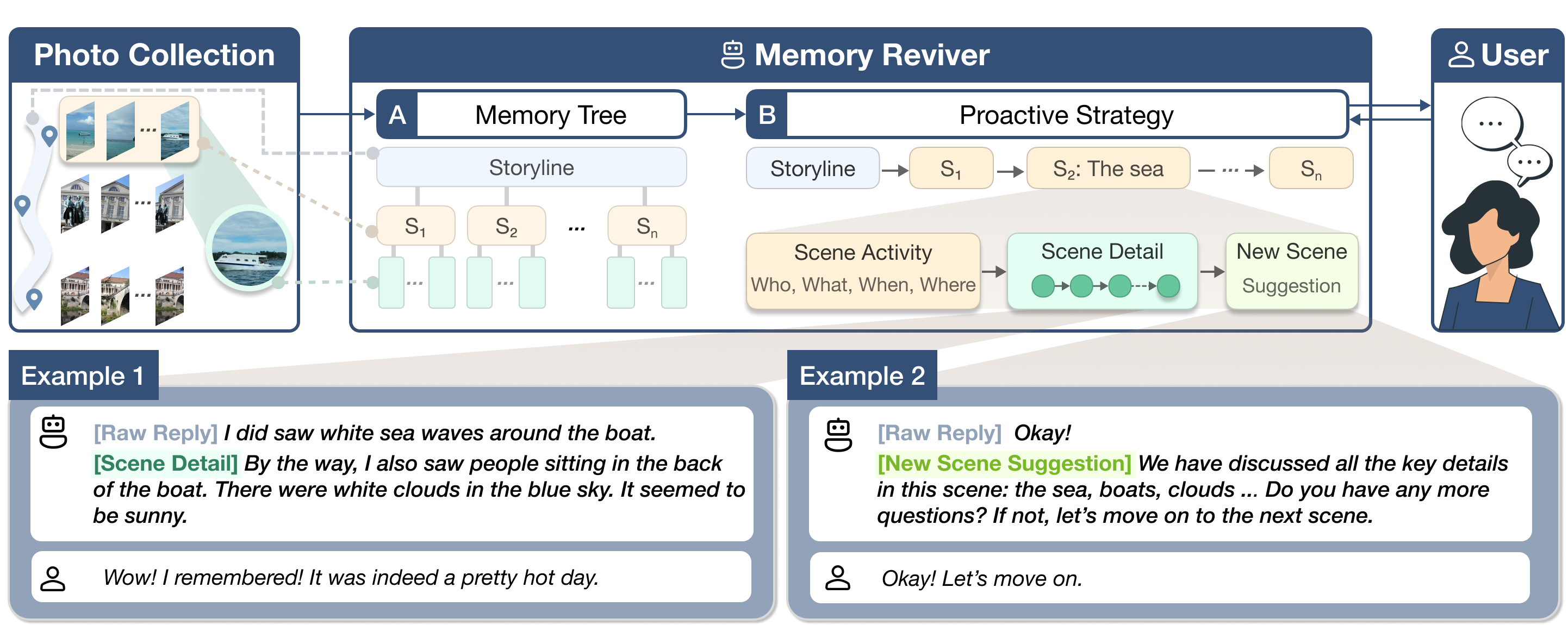}
  \caption{
  Memory Reviver is a proactive chatbot that actively guides people with visual impairment (PVI) to reminisce with a photo collection. It incorporates two novel features: (A) a \emph{Memory Tree}, which uses a hierarchical structure to organize information in a photo collection; and (B) a \emph{Proactive Strategy}, which actively delivers information to users at proper conversation rounds. Powered by the two features, Memory Reviver begins the chat with a clear storyline, helps users recall past activities, enriches their memories by gradually presenting details (Example 1), and suggests new scenes at proper rounds (Example 2).
  }
  \label{fig:teaser}
\end{teaserfigure}

%\settopmatter{printfolios=true}
\maketitle

\section{Introduction}
Reminiscence, the activity of recalling life experiences from one's past \cite{webster2010mapping, white2023memory}, has been shown to improve mental well-being \cite{bryant2005using, lyubomirsky2005pursuing}, foster social connections \cite{allen2009legacy}, and promote personal growth \cite{crete2012reconstructing,fivush2011making}. 
This process allows individuals to reflect on memorable moments and milestones, often by browsing through photo collections \cite{chen_exploring_2023,axtell_design_2022,mcgookin_reveal_2019}. 
However, reminiscing with a photo collection is challenging for people with visual impairment (PVI). 
Although tools have been explored to help PVI examine individual photos \cite{stangl_going_2021,stangl_person_2020,jung_so_2022,nair_imageassist_2023}, they lack features for a comprehensive reminiscing experience, such as providing a holistic overview of the photo collection. 
Consequently, PVI often rely on sighted people to reminisce with photo collections \cite{yoo_understanding_2021,jung_so_2022}. 
This dependence limits the frequency of their reminiscing activities and impedes them from deriving the associated psychological benefits \cite{yoo_understanding_2021,jung_so_2022}.

To address this limitation, we propose designing a chatbot that enables PVI to reminisce with photo collections through natural language communication. This chat-based approach is already familiar to PVI through their interactions with sighted helpers \cite{bigham2010vizwiz}, rendering it a suitable alternative when sighted help is unavailable or not preferred \cite{jung_so_2022}. Moreover, chatbots have been shown to effectively support emotional communication \cite{seo2023chacha,wei2023leveraging}, making them a promising tool for facilitating reminiscence.

To inform the chatbot design, we conducted a formative study with eight PVI. 
They were invited to reminisce with their photo collections by conversing with a naïve chatbot based on GPT-4V \cite{yang2023gpt4v}, the state-of-the-art large multimodal model. 
Our research uncovered significant shortcomings in directly utilizing GPT-4V for reminiscence activities among PVI, primarily due to disorganized conversation flow and the chatbot's non-proactive interaction style. 
Firstly, the ``one question, one answer'' communication style led to the information being scattered across multiple rounds. This made it hard for participants to recall and organize details, preventing them from forming a clear story about their past. 
Secondly, since participants could not visually explore new scenes, the chatbot's lack of proactive guidance further challenged participants to engage deeply in the conversation.

To address these challenges, we present Memory Reviver, a proactive chatbot that actively guides PVI to reminisce with a photo collection. Memory Reviver is tailored for photo collections about a specific event. It incorporates two novel features: (1) an information architecture of \textbf{\emph{Memory Tree}}, which extracts the information in a photo collection into a hierarchical structure; and (2) a \textbf{\emph{Proactive Strategy}}, which actively delivers information to users at recognized proper conversation rounds. 
Powered by the two features, Memory Reviver delivers a natural conversation flow. It begins the conversation with a clear storyline, helps users recall past activities, and enriches their memories by gradually presenting details. After thoroughly examining the details of a specific scene, Memory Reviver proactively suggests new scenes. This seamless blend of guided exploration with free-form dialogue allows Memory Reviver to offer an experience that closely resembles natural reminiscence.

To evaluate Memory Reviver, we conducted a within-subject study with 12 PVI. The participants reminisced with two personal photo collections using Memory Reviver and the naïve GPT-4V chatbot (baseline), respectively. 
Results showed that Memory Reviver enabled the exploration of significantly more scenes ($p<.01$) and helped users recall more memories ($p<.01$) than the baseline. 
Subjective ratings showed that Reviver outperformed the baseline in facilitating engaging reminiscence, aiding in photo collection understanding, and delivering natural conversational experiences. 
Moreover, participants reported experiencing strong positive emotions and self-reflection using Memory Reviver. 
We further discussed directions for personalizing Memory Reviver and summarized implications that could inform future designs of photo reminiscence support and accessible chatbots for PVI.

Our contributions are threefold:
\begin{itemize}
\item We present Memory Reviver, a proactive chatbot that leverages two novel features—a \emph{Memory Tree} and a \emph{Proactive Strategy}—to offer guided reminiscence experiences for PVI;
\item We contribute an evaluation study that demonstrates how PVI engage with reminiscence using Memory Reviver;
\item We distill implications that could guide future designs of photo reminiscence support and accessible chatbots for PVI.
\end{itemize}
\section{Related Work}
Our work extends prior research in three areas: (1) photo reminiscence support, (2) image understanding support, and (3) chatbot design for PVI.

\subsection{Photo Reminiscence Support for PVI}
Reminiscing with photos offers significant psychological benefits, enabling individuals to reflect on the past \cite{chen_exploring_2023,jones_co-constructing_2018} and envision the future \cite{mcgookin_reveal_2019,axtell_design_2022}. Prior research has highlighted the desire of PVI to engage in photo reminiscence \cite{zhao2017effect,harada2013accessible,jung_so_2022,yoo_understanding_2021}.

To support PVI in reminiscing with individual photos, prior works have explored various tools, including AI-generated image descriptions \cite{zhao2017effect, jung_so_2022}, audio recordings \cite{harada2013accessible}, and tactile photography \cite{yoo_understanding_2021}. For instance, Jung et al. \cite{jung_so_2022} investigated the effectiveness of current AI-generated image descriptions for reminiscence and found that these descriptions often focus solely on visual elements (e.g., ``\textit{A plate of food on a table}''), making it challenging for PVI to recall associated memories. To address this limitation, Harada et al. \cite{harada2013accessible} proposed linking photos with audio recordings of past moments to assist PVI in reminiscing. Additionally, Yoo et al. \cite{yoo_understanding_2021} suggested using tactile photography to facilitate PVI's engagement with photos. However, these approaches primarily focused on single-photo reminiscence, requiring users to examine each photo individually and manually organize large amounts of information, resulting in a tedious experience.

Recent advances in multi-image-to-text generation techniques have the potential to enable PVI to comprehend multiple images as a whole. These techniques included multi-image summarization \cite{yang_knowledgeable_2019, yu_transitional_2021, jung_hide-and-tell_2020, huang_visual_2016} and multi-image visual question answering \cite{BLIP1, BLIP2, penamakuri2023answer}. Recent advances in large multi-modal models (LMM) offer promising opportunities for PVI to comprehend photo collections through natural language input \cite{yang2023gpt4v,team2023gemini}. However, there is a lack of understanding of how these techniques could be leveraged to support their needs. We take an initial step to examine this problem where we invited PVI to reminisce with a photo collection through interacting with a state-of-the-art LMM. Our study uncovered four types of information that PVI seek during reminiscence: a storyline, scene activities, scene details, and new scenes. We further provide practical guidelines on how to distill such information leveraging automated techniques.

\subsection{Image Understanding Support for PVI}
Enabling image access for PVI has been a consistent research focus in HCI. PVI primarily access images through image descriptions created by manual \cite{bigham2010vizwiz} and automatic methods \cite{vinyals2015show, xu2015show}. Recent studies have investigated how PVI use image descriptions to access online images \cite{zhao2017effect, wu2017automatic} and how their information needs vary across different scenarios \cite{jung_so_2022, stangl_going_2021, stangl_person_2020, morris_rich_2018}. For example, Stangl et al. \cite{stangl_person_2020} found that PVI's preferences for image descriptions vary depending on the image source and the surrounding context.

To fulfill the diverse user needs, several works have explored methods for interactive image exploration, including visual question answering \cite{bigham2010vizwiz, gurari2019vizwiz, vqa, brady2013visual} and touch-based image exploration \cite{nair_imageassist_2023, lee2021image, lee2022imageexplorer, zhong2015regionspeak}. For example, ImageAssist \cite{huh2022cocomix} designed three tools to help PVI explore different regions of interest in an image. However, these works mainly focus on enhancing the exploration of individual images, which is inadequate for complex visual content, such as multiple images \cite{huh2023genassist}, multi-page slides \cite{peng2023slide}, and multi-shot videos \cite{van2024making}. Compared to a single image, these complex forms present challenges due to excessive information \cite{huh2023genassist}. To address this challenge, recent research \cite{huh2023genassist, peng2023slide, van2024making} has explored methods to present information associated with complex visual content. For example, GenAssist \cite{huh2023genassist} summarized the similarities and differences among multiple images into an easy-to-compare table format. ShortScribe \cite{van2024making} organized information from short-form videos into hierarchical summaries. However, most existing works focus on presenting information via screen readers, and few works have explored how to convey massive visual information through chat-based interaction. Our work builds on this literature by examining what information in a photo collection facilitates reminiscence (i.e., Memory Tree) and how to present this information via chatbots.

\subsection{Chatbot Design for PVI}
Chatbots, also known as conversational agents and dialog systems, are software systems designed to simulate human-like conversations through the analysis of natural language data \cite{dialougue_system}. Due to their conversational nature, chatbots have emerged as promising tools to facilitate accessible interaction for PVI \cite{pucci2023defining, choi2020nobody}. For example, Pucci et al. demonstrated that chatbots can offer a more natural interaction for web browsing in comparison to screen readers, allowing PVI to directly express their intentions for information retrieval \cite{pucci2023defining}. To inform the design of chatbots for PVI, previous research has proposed general design frameworks \cite{lister2020accessible,neil2014mobile,stanley2022chatbot}. Recent studies have investigated the accessibility issues of voice assistants like Siri \cite{branham_reading_2019, pradhan_accessibility_2018, abdolrahmani_siri_2018,choi2020nobody}. For example, Pradhan et al. \cite{pradhan_accessibility_2018} found that some PVI struggled to discover the commands supported by a chatbot. However, existing works have primarily focused on designing chatbots to execute specific commands for PVI (e.g., controlling household appliances), leaving a notable gap in the design of chatbots aimed at engaging PVI in reminiscing with photos.

To bridge this gap, we closely involved PVI in our design process. Our findings reveal that the passive ``one question, one answer'' communication style challenges PVI to engage in the conversation and construct a clear personal story. We thus organize the information needed during reminiscence into a hierarchical structure and offer proactive guidance.

\section{Formative Study}
To understand the needs and challenges of PVI in reminiscing with a photo collection, we conducted a formative study with eight PVI. The formative study comprised (1) a semi-structured interview to understand their current practices and challenges, and (2) a tech-probe session to investigate their needs and challenges related to reminiscing with a chatbot.

\subsection{Methods}
\subsubsection*{\textbf{Participants}}
We recruited eight PVI (P1-P8; four males, four females; Table~\ref{tab:demographics} lists demographics) from an online support community. Their ages ranged from 26 to 45 (mean = 32.9, SD = 6.4). Five participants were totally blind and three participants were legally blind with light perception. All participants had previous experience using image accessibility tools (e.g., \textit{Be My AI} \cite{be_my_eyes} and screen readers) and chatbots (e.g., ChatGPT and Siri). They regularly reminisced with photos and consented to share photos for study purposes. All participants were native Mandarin speakers. \footnote{In the paper, participants' original speech was translated from Mandarin into English.}.

\subsubsection*{\textbf{Photos}} Each participant provided a photo collection related to a specific event, which can be a trip, a gathering, or a performance. Each collection contained 21 to 57 photos (mean = 39.9).

\subsubsection*{\textbf{Procedure}}
The study consisted of two successive phases: a 0.5-hour interview followed by a 1.5-hour tech-probe session, with a break of 5 minutes in between.

\textbf{Phase 1: Semi-structured Interview}. During the interview, we asked participants about their current practices and challenges via the following questions: (1) What's your current practice in reminiscing with photo collections? (2) What are the challenges encountered? (3) What are your expectations for reminiscing with photo collections? To help participants recall previous experiences, we asked about concrete situations, such as ``\textit{Can you recall the last time you reminisced with photos?}''. When interesting points came up, we followed up with questions to probe additional details. 

\textbf{Phase 2: Tech-probe Session}. In this session, participants first received a brief tutorial on a tech-probe chatbot and then used it to reminisce with their photo collections. They conducted free-form conversations with the chatbot. 
After the session, we conducted an exit interview to understand the participants' needs and challenges in reminiscing with the chatbot. To help them recall their needs and challenges, we read their chat histories with the chatbot aloud to them. The whole study was conducted via a one-on-one Zoom session. Participants were compensated 160 CNY for their time.

\subsubsection*{\textbf{Tech-probe: a naïve chatbot}}
\begin{figure}[!htbp]
    \centering
    \includegraphics[width=\linewidth]{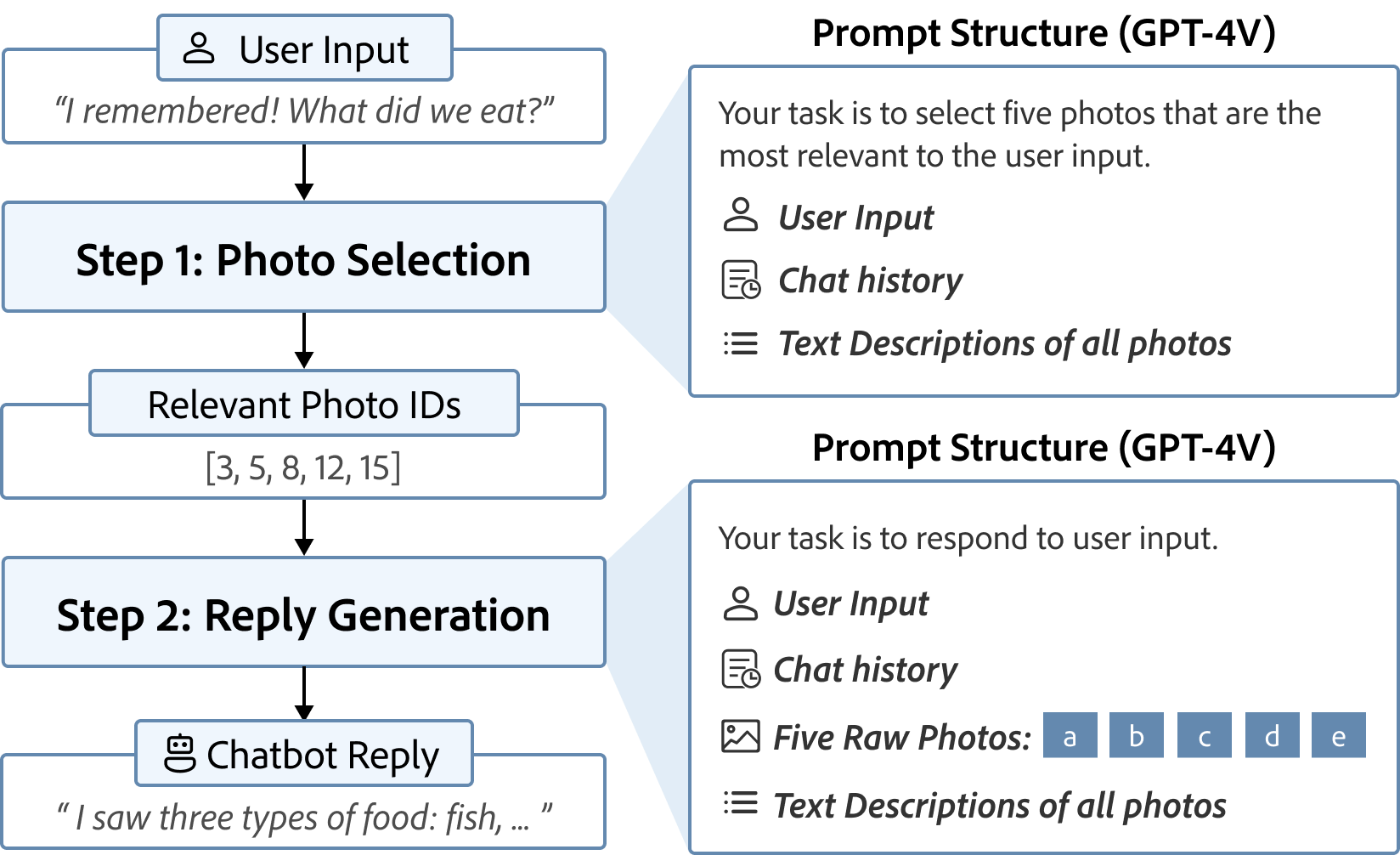}
    \caption{The pipeline of the naïve chatbot.}
    \label{fig:initial}
\end{figure}

We developed a naïve chatbot for the tech-probe session. The chatbot was designed to converse with users based on their utterances and their photo collections. To enable the chatbot to access photo collections, we used GPT-4V to generate natural-language replies based on image inputs \cite{yang2023gpt4v}.

Each time the user provides an input sentence, the system selects the five most relevant photos to generate a reply. Specifically, the system employs a two-step pipeline (see Figure~\ref{fig:initial}). In the first step, it selects five photos by prompting\footnote{The prompts are provided in the supplementary materials.} GPT-4V with the text descriptions of all photos. These text descriptions are respectively pre-generated by GPT-4V for each image. Then, in the second step, the system passes the raw data of the five selected photos into GPT-4V to generate a final reply. This pipeline allows the chatbot to access photos in the collection while managing the length of input prompts. We did not directly prompt GPT-4V with the raw images of the entire photo collection due to token limits \cite{yang2023gpt4v} and the negative impact of long input prompts on performance \cite{wu2022ai,wei2023leveraging,brown2020language}. Additionally, to enable the chatbot to recognize the user in the photo collection, an extra portrait photo of the user is inputted into GPT-4V.

Users interacted with the naïve chatbot through verbal communication in Mandarin. The chatbot used Amazon Transcribe \footnote{https://aws.amazon.com/pm/transcribe/} for speech recognition and Amazon Polly \footnote{https://aws.amazon.com/polly/} for output speech synthesis. The chatbot operated on the experimenter's desktop and received participants' verbal input directly via Zoom. The average response time\footnote{This response time refers to the period after the user finishes speaking and before the chatbot starts replying.} of the chatbot was 16.7 seconds (SD=5.3, MAX=31.2), which was reported to be acceptable by the participants.

\subsubsection*{\textbf{Analysis}}
We recorded the study audio, transcribed it, and then qualitatively coded it based on (1) participants' current practice and challenges; (2) their needs and challenges in reminiscing with the chatbot. 
Additionally, participants' inputs during their conversations with the chatbot were categorized according to their information needs. Two authors conducted the open coding process separately and reached an agreement through discussions. Based on the above data, we report our findings.

\subsection{Interview Findings}\label{sec:formative_goals}

The semi-structured interview revealed participants' current practices, challenges, and goals during reminiscence. 

\subsubsection*{\textbf{Current Practices and Challenges}} 

Participants reported saving photo collections for various events, including trips, ceremonies, and stage performances. To make reminiscence accessible, participants adopted common photo-organizing practices. Seven out of eight participants indicated that they intentionally organized photos for each memorable event into separate albums. One participant (P4) mentioned using time frames (e.g., New Year's Day) to filter collections related to specific events.

Despite their efforts in organizing photos, participants faced challenges when reminiscing with photo collections that were ``\textit{captured with sighted help}'' (P3) and those that were ``\textit{taken many years ago}'' (P1). The challenges stemmed from their \textbf{limited memories} about the photos. Due to their limited memories, participants noted that reminiscing with such a photo collection is ``\textit{unachievable without sighted help}'' (P3). However, even with sighted help, their needs were often unmet because ``\textit{sighted people rarely have the patience to describe the details.}'' (P1). Additionally, all participants tried to use AI-image descriptions to reminisce with a photo collection, but they found these descriptions ``\textit{too simple to trigger memories}'' (P8) and the process too tedious (P2: ``\textit{It's tedious to examine each photo one by one, especially when many photos are similar.}''). Consequently, all participants expressed that the current methods were far from effective in supporting reminiscence.

\subsubsection*{\textbf{Two common goals.}}
Participants shared two common goals for reminiscence. First, all eight participants expressed a desire to \textbf{relive all the scenes} in a photo collection. For instance, P3 expressed, ``\textit{So many precious scenes were captured during my wedding. I don't want to miss out on any scenes.}''. Moreover, five participants hoped to \textbf{recall as many memories as possible}. As P2 noted, ``\textit{I hope my past memories can be revived to the fullest.}''. These two goals informed our task metrics in the evaluation study.

\subsection{Tech-probe Findings}\label{sec:formative_findings}
The tech-probe session unveiled participants' needs and challenges in reminiscing with a chatbot. 

\subsubsection*{\textbf{Four Types of Information Needs}}
When reminiscing with the chatbot, participants commonly reminisced on a scene basis. For example, P7 noted, ``\textit{I felt like I was moving through different scenes. I found a scene, explored it to the fullest... and moved on to a new one.}''. During the conversations, participants generated a total of 171 inputs. These inputs were categorized into four types\footnote{There were 14 inputs classified under ``others'' which included questions about retouching photos and text-to-image generation.} based on their information needs: a storyline (13), scene activities (45), scene details (71), and new scenes (28). The needs associated with each information type are as follows:

\textbf{1. A Storyline}: All participants requested the chatbot to generate a storyline encompassing all the scenes in a photo collection. For example, P3 asked the chatbot, ``\textit{Could you summarize the whole story in the collection?}'' However, participants found that the storyline provided by the chatbot lacked a clear and chronological structure. For instance, P5 mentioned, ``\textit{It mixed up many scenes together... and it did not follow a chronological order.}'' This finding highlights that \textbf{the storyline should be structured as a series of scenes in chronological order (D1)}.

\textbf{2. Scene Activities}: When participants focused on a scene of interest (e.g., ``\textit{by the sea}''), they typically started by gathering clues to help them recall past activities: ``\textit{recalling what I did is the basis of reminiscence, otherwise photos are just public images to me.}'' (P5).

When recalling past activities, participants mainly asked questions regarding ``\textit{who, what, when, and where}'' (4W) aspects. For example, P3 progressively asked ``\textit{Was I alone by the sea? (who)}'', ``\textit{What was I doing? (what)}'', and ``\textit{Was it daytime or night? (when)}''. Table~\ref{tab:space_activity} summarizes the strategies used by participants to uncover the 4W aspects. \textbf{One common strategy is to ask about texts in the photos} because ``\textit{many texts contain time and location details, such as entrance signs and holiday banners.}'' (P1).

However, participants found asking such questions tedious, because ``\textit{I didn't know what clues were present in the photos, so I had to try many times.}'' (P3). Additionally, participants experienced feelings of doubt and confusion when the reasons for the 4W aspects were not explicitly stated. For example, P2 noted ``\textit{I'm confused why the chatbot said it looked like winter. Is it because of our clothing? I need to know the reasons.}''. Collectively, findings show the importance of \textbf{helping users recall past activities by presenting the ``who, what, when, and where'' aspects with reasons (D2)}.

\textbf{3. Scene Details}: After recalling past activities, participants proceeded to inquire the chatbot about visual details, such as colors and shapes of the objects existing in the scene. Table~\ref{tab:space_details} summarizes the details inquired by participants, which align with findings in prior works \cite{stangl_going_2021,stangl_person_2020}. Their expectations were to enrich their understanding of past experiences: ``\textit{All the details reconnected me with the moment: people's poses, their facial expressions... They enrich my memories.}'' (P6). 

However, they found it hard to fully explore a scene by asking questions: ``\textit{Since I can only ask details about objects I already know, I always fear that I have missed out something interesting.}'' (P1). While participants tried to ask for a list of details, they found the reply to be overwhelming: ``\textit{It is hard to grasp so many details at one time.}'' (P4). Collectively, these findings highlight the tension between the need for in-depth exploration and the difficulty in achieving it by asking questions. Thus, it is important for the chatbot to \textbf{progressively present scene details to help users fully explore a scene (D3)}.

\textbf{4. New Scenes}: After exploring one scene, participants would find a new scene of interest, so as to ``\textit{fully relive all the scenes}'' (P5). However, since participants couldn't visually discover new scenes, they relied on the chatbot to provide guidance. For example, P3 asked four times, ``\textit{Any other scenes?}''. However, participants found that the naïve chatbot either replied ``\textit{a random scene}'' (P6) or ``\textit{a scene that has been discussed}'' (P3), leading to confusion about the sequence of scenes and how many were left to explore. This issue stemmed from the memory management of large multi-modal models, a common problem reported in prior works \cite{zhong2024memorybank,wang2024augmenting}. This finding highlights the importance for the chatbot to \textbf{actively suggest new scenes at proper conversation rounds and clearly inform users of the discussion progress (D4).}

\begin{table}[!hbtp]
\centering
\caption{Strategies used by PVI to recall scene activities.}
\renewcommand\arraystretch{1.1}
\resizebox{1\columnwidth}{!}{
\begin{tabular}{c|l}
\hline
\textbf{Aspects} & \textbf{Strategies to recall scene activities from photos} \\
\hline
\multirow{3}{*}{Where} & \textbf{Landmarks}: e.g., the Eiffel Tower. \\
 & \textbf{Surroundings}: the sea, hills, canteens, museums, etc. \\
 & \textbf{Places in Texts}: entrance signs, holiday banners, etc. \\
\hline
\multirow{3}{*}{When} & \textbf{Season}: clothes, tree leaves, etc. \\
 & \textbf{Day or Night}: lighting conditions, etc. \\
 & \textbf{Time in Texts}: holiday banners, screens, etc. \\
\hline
\multirow{2}{*}{Who} & \textbf{Visual Appearances}: age, gender, clothes, hair, etc. \\
 & \textbf{Names in Texts}: name tags, name badges, etc. \\
\hline
\multirow{3}{*}{What} & \textbf{Human Actions}: e.g., playing musical instruments. \\
 & \textbf{Objects}: food, animals, roller coasters, etc. \\
 & \textbf{Actions in Texts}: menu, conference banners, etc.\\
\hline
\end{tabular}
}
\label{tab:space_activity}
\end{table}

\begin{table}[hbtp]
\centering
\caption{Scene details asked by PVI.}
\renewcommand\arraystretch{1.1}
\resizebox{1\columnwidth}{!}{
\begin{tabular}{c|l}
\hline
\textbf{Aspects} & \textbf{Scene Details asked by PVI} \\
\hline
\multirow{2}{*}{People} & Number of people in the photo \\
 & Gender, age, hair, clothes, facial expression, pose, etc. \\
\hline
\multirow{1}{*}{Food} & Name, color, shape, etc. \\
\hline
\multirow{1}{*}{Animals} & Breed, color, size, etc. \\
\hline
\multirow{1}{*}{Plants} & Species, color, shape, height, etc. \\
\hline
\multirow{1}{*}{Buildings} & Color, shape, style, etc. \\
\hline
\multirow{1}{*}{Texts} & Raw text, position of the text (e.g., on a screen) \\
\hline
\multirow{1}{*}{Others} & Color, shape, etc. \\
\hline
\end{tabular}
}
\label{tab:space_details}
\end{table}

\subsubsection*{\textbf{Two Key Challenges}}

We identify two key challenges faced by PVI when using the naïve chatbot to reminisce.

\textbf{The first challenge is the scattering of information} throughout the conversation. While participants actively sought the four types of information (D1-D4), the ``one question, one answer'' communication style led to information being scattered across multiple rounds, making it tedious for participants to recall and organize details. Moreover, the conversation flow also lacks organization. This was evident as the storyline was unstructured and the order of scenes was random and repetitive. This highlights the need for organizing the information into a clear structure.

\textbf{The second challenge is a lack of proactive guidance} from the chatbot. Since participants couldn't visually explore new scenes, the chatbot's lack of proactive guidance further challenged participants to engage in the conversation. As P1 noted, ``\textit{Friends would naturally introduce new topics. But with this chatbot, I had to find something new by myself. It didn't feel natural.}'' This highlights the importance of the chatbot offering proactive guidance to help users explore a photo collection.

Based on the identified information needs and challenges, we distill two implications for a photo reminiscence chatbot for PVI: (1) \textbf{Clear Information Organization}: organize the four types of information into a clear structure. (2) \textbf{Proactive Guidance}: proactively lead the conversation and provide information to users at proper conversation rounds. These implications motivated the design of Memory Reviver.
\section{Memory Reviver}
We present Memory Reviver, a proactive chatbot designed to actively guide users in reliving all scenes within a photo collection. It addresses the two challenges through two novel features: (1) an information architecture of \textbf{\emph{Memory Tree}}, which organizes the information in a photo collection into a hierarchical structure; and 
(2) a \textbf{\emph{Proactive Strategy}}, which actively guides users to explore a photo collection. Together, these features help users gain a clear and thorough understanding of a photo collection during reminiscence.

\subsection{Memory Tree}
\emph{Memory Tree} aims to organize information in a photo collection into a structured format. Our formative study reveals four types of information essential for reminiscence: (1) a storyline, (2) scene activities, (3) scene details, and (4) new scenes. Inspired by the organization of autobiographical memory \cite{conway1987organization},  we designed the \emph{Memory Tree} as a three-level tree structure (see Figure~\ref{fig:memory_tree} (b)).

\subsubsection*{\textbf{The Structure of Memory Tree}}
\begin{figure*}[!htbp]
    \centering
    \includegraphics[width=1\linewidth]{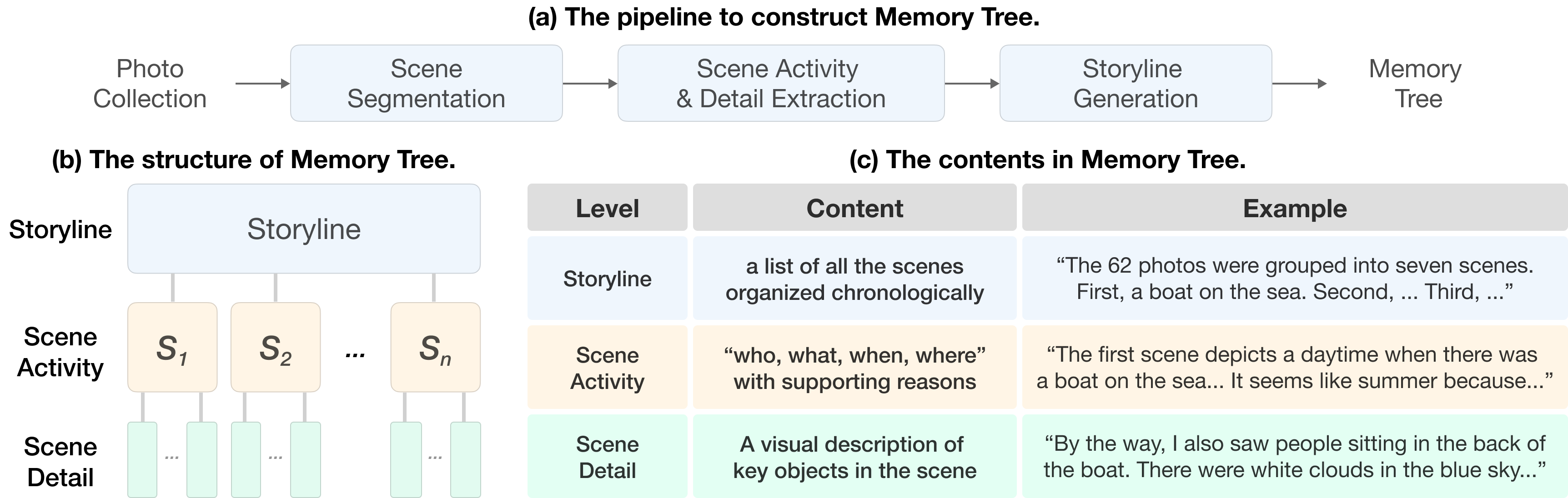}
    \caption{The \emph{Memory Tree} organizes information in a photo
    collection into a three-level structure.% (a) The pipeline to construct Memory Tree. (b) The structure of Memory Tree. (c) The contents in Memory Tree.
    }
    \label{fig:memory_tree}
\end{figure*}

The \emph{Memory Tree} has three levels: ``storyline - scene activity - scene detail''. The first level is the \textbf{storyline (D1)}, which contains a list of all the scenes arranged chronologically. The second level is the \textbf{scene activity (D2)}, which organizes the ``who, what, when, and where'' aspects into a sentence. These aspects are extracted using the user strategies in Table~\ref{tab:space_activity}. The third level is the \textbf{scene detail (D3)}. Each detail comprises a visual description of objects in the scene (e.g., people, food, animals, etc.), which is extracted according to the guidelines outlined in Table~\ref{tab:space_details}. \textbf{New scenes (D4)} can be directly retrieved from the storyline. The contents and examples in each level are shown in Figure~\ref{fig:memory_tree} (c).

\begin{figure*}[!htbp]
    \centering
    \includegraphics[width=1\linewidth]{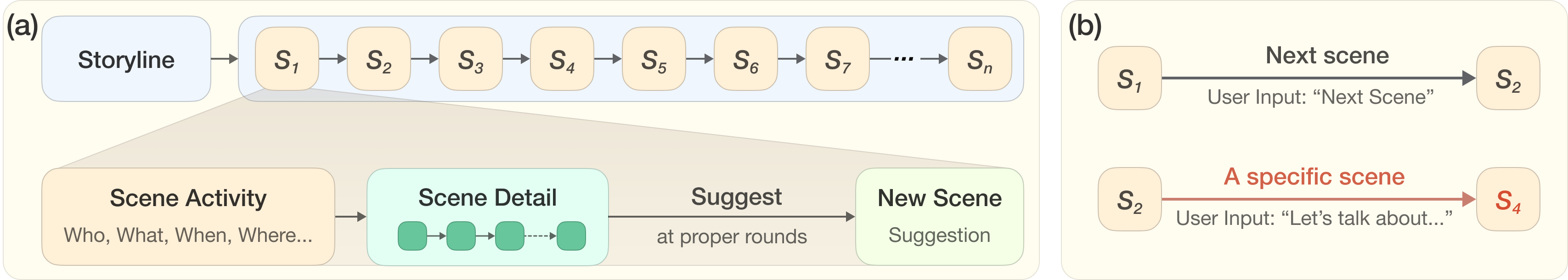}
    \caption{(a) The \emph{Proactive Strategy} starts the conversation with a storyline and then guides users to relive each scene. Within each scene, it introduces the scene activity, progressively presents scene details, and suggests the next scene at proper conversation rounds. (b) Users can freely switch scenes using two natural language commands.}
    \label{fig:4_strategy}
\end{figure*}

\begin{figure*}[!htbp]
    \centering
    \includegraphics[width=1\linewidth]{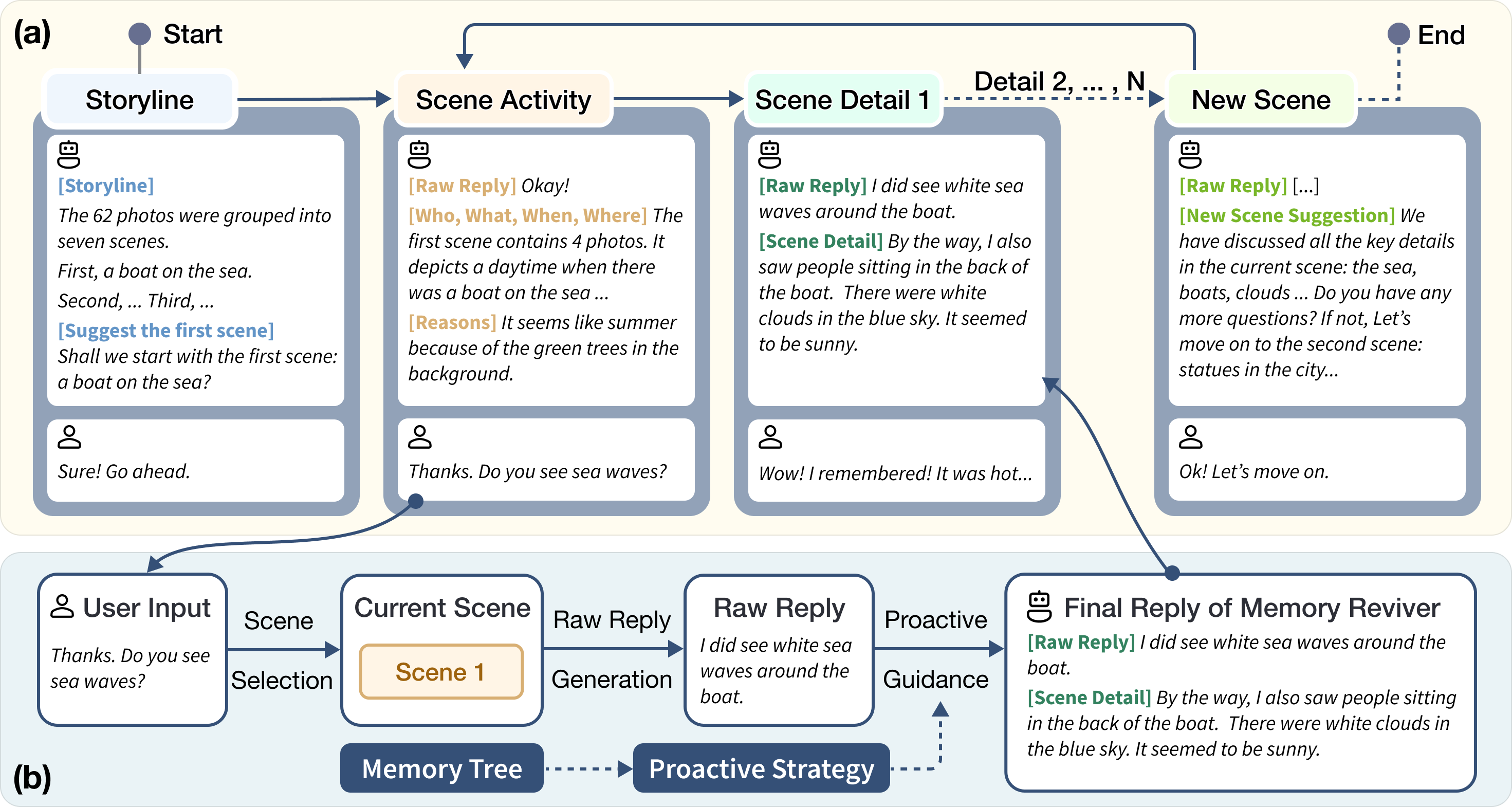}
    \caption{(a) Examples of the multi-round conversations. (b) The pipeline for Memory Reviver to generate a reply in each round.}
    \label{fig:4_walkthrough}
\end{figure*}

\subsubsection*{\textbf{Constructing the Memory Tree}}
Memory Reviver is designed to handle a photo collection about a specific event. After the users upload such a photo collection, the system first arranges the photos in chronological order using the original timestamps in the metadata. It then constructs the \emph{Memory Tree} by segmenting the collection into scenes and distilling information for each scene. To achieve this goal, we leverage the capabilities of GPT-4V \cite{yang2023gpt4v} in recognizing scenes, activities, landmarks, faces, and texts. We use a three-step pipeline to construct the \emph{Memory Tree} (see Figure~\ref{fig:memory_tree} (a)):

\textbf{(1) Scene Segmentation}: In our formative study, users perceived a scene to consist of consecutive photos capturing the same activity (``\textit{who, what, when, where}''). Leveraging this insight, we employ GPT-4V to assess the activity similarity between two adjacent photos on a scale of 0 to 1. Empirically, we determine that a segmentation point occurs when the rating is below 0.5. To be noted, this module is not claimed as a contribution in our work and can be substituted with advancements in computer vision \cite{garcia2018predicting,aakur2019perceptual,du2022fast}.

\textbf{(2) Scene Activity and Detail Extraction}: After segmenting the collection into the scenes, we extract information from each scene by inputting the photos into GPT-4V. We prompt\footnote{The prompts are provided in the supplementary materials.} the model with the guidelines outlined in Table~\ref{tab:space_activity} and Table~\ref{tab:space_details}. To ensure the comprehensiveness of the extracted information, all photos from each scene are input together into GPT-4V. Additionally, a portrait photo of the user is inputted for user recognition.

\textbf{(3) Storyline Generation}: After extracting the information for all scenes, we use GPT-4V to generate the storyline. The GPT-4V is prompted with the information of all scenes and the task description (i.e., ``\textit{Summarize each scene briefly with a short sentence and then list them in chronological order from the beginning to the end.}'')

The pre-extracted information in the \emph{Memory Tree} is then used by the \emph{Proactive Strategy} to guide the conversation.

\subsection{Proactive Strategy}
The \emph{Proactive Strategy} delivers the information in the \emph{Memory Tree} at proper conversation rounds, forming a natural conversation flow. In achieving this, it employs a state machine \cite{seo2023chacha, winograd1986language} to guide the conversation. The state machine is shown in Figure~\ref{fig:4_strategy} (a).

\subsubsection*{\textbf{Proactive Strategy Design}}
The \emph{Proactive Strategy} starts the conversation with a storyline and then guides users to relive each scene. Within each scene, it introduces the scene activity, progressively presents scene details, and suggests the next scene at proper conversation rounds. The specific conversation flow is as follows:

\textbf{1. Start the conversation with a storyline}: To help users quickly skim through the collection, Memory Reviver starts with a storyline (see Figure~\ref{fig:4_walkthrough} (a)). The storyline lists all the scenes chronologically to match users' past experiences \textbf{(D1)}. After introducing the storyline, Memory Reviver would suggest starting the exploration of the first scene in the photo collection.

\textbf{2. Introduce each scene with scene activities}: When users enter each scene for the first time, the chatbot will present the scene activity (see Figure~\ref{fig:4_walkthrough} (a)). The scene activity encompasses the ``who, what, when, and where'' aspects \textbf{(D2)}. Explanations are provided on how these aspects are determined from photos (e.g., ``\textit{It looked like a canteen because of the 'student canteen' sign at the entrance.}''). The whole sentence does not exceed 100 characters in length.% retrieved from Memory Tree 

\textbf{3. Present scene details gradually}:
After the scene introduction, the chatbot will progressively present scene details \textbf{(D3)}. In each round of reply, the chatbot will scan through the list of scene details in the \emph{Memory Tree} and present the first detail that has not been discussed in the previous conversation.

\textbf{4. Suggest new scenes at proper rounds}:
To help users explore all the scenes, Memory Reviver would actively suggest a new scene at recognized proper conversation rounds \textbf{(D4)}. A proper round is defined as when users no longer show interest in the current scene. We adopt the following \textbf{criteria to detect this round}: all the scene details in the current scene have been discussed and users have no questions in the last round. Upon detecting this round, Memory Reviver will scan through the storyline and suggest the first scene that has not yet been discussed. This design utlizes the psychological findings that forward recall offers the fastest access to past memories \cite{conway2019structure}. The scene suggestion is accompanied by a summary of the current scene, which aims to address any remaining questions users may have before switching to the new scene.

\textbf{5. Allow free scene switch}: Apart from actively suggesting the next scene by default, Memory Reviver allows users to flexibly switch among scenes using natural-language commands. Users can employ ``\textit{Next scene}'' to advance to the next scene or ``\textit{Let's talk about ...}'' to switch to a specific scene (see Figure~\ref{fig:4_strategy} (b)).

When all scenes have been discussed, Memory Reviver concludes by providing a summary. 
This summary serves to help users construct a clear story of their past.

\subsubsection*{\textbf{Integrating the Proactive Strategy into Free-form Dialogues}}
The above design of the \emph{Proactive Strategy} is integrated into free-form conversations, to prevent unnatural interactions — a common challenge in proactive chatbot design \cite{tallyn2018ethnobot,chabot_interact}. To achieve this goal, Memory Reviver first responds to the user input. Then, it provides proactive guidance by using the \emph{Proactive Strategy} to retrieve information from the \emph{Memory Tree}.

Each time the user speaks an input sentence, Memory Reviver employs a three-stage pipeline to generate a reply (see Figure~\ref{fig:4_walkthrough} (b)). The pipeline first (1) identifies a specific scene related to the user input, then (2) generates a raw reply, and finally (3) integrates the proactive guidance into the reply.

\textbf{(1) Scene Selection}: Upon receiving a user input, Memory Reviver determines the current scene using the following rules: (a) If Memory Reviver has suggested a new scene in the last round (including suggesting the first scene at the beginning), it checks whether the user accepts the suggestion by detecting keywords (e.g., ``\textit{Okay}'', ``\textit{Go on}'', etc.); (b) If the user input contains ``\textit{Next Scene}'', Memory Reviver moves directly to the next scene; (c) If the user input contains ``\textit{Let's talk about [keyword]}'', Memory Reviver matches the keyword with the scene details in each scene to identify a specific scene; (d) Otherwise, the scene remains unchanged from the last time (the scene is initially set to scene 1).

The above rules limit Memory Reviver to provide replies on a single-scene basis. We discussed the necessity for addressing cross-scene questions in section \ref{sec:concerns_improvements} of the evaluation study.

\textbf{(2) Raw Reply Generation}: Memory Reviver uses GPT-4V to generate a raw response. The GPT-4V is prompted\footnote{The prompts are provided in the supplementary materials.} with the task description ``\textit{Your task is to generate a response to the user input, based on the chat history and the photos.}'', along with the user input, the chat history, and the raw photos in the current scene.

\textbf{(3) Proactive Guidance}: 
Ultimately, Memory Reviver combines proactive guidance with the raw reply. It first retrieves the proactive guidance from the \emph{Memory Tree} using the \emph{Proactive Strategy} outlined in Figure~\ref{fig:4_strategy} (a). Then, it concatenates the raw reply with the proactive guidance to form a final response. This final reply is subsequently provided to users.

\subsection{Implementation}
Memory Reviver was implemented as a Python desktop program, with the \emph{Memory Tree} stored in a JSON file and the \emph{Proactive Strategy} implemented as a rule-based state machine \cite{seo2023chacha}. For the AI model, we used gpt-4-vision-preview with a temperature value of 0.8. Users interacted with Memory Reviver through verbal communication in Mandarin. We implemented the voice interaction of Memory Reviver using Amazon Transcribe to recognize users' input speech and Amazon Polly to synthesize output speech.
\section{Technical Evaluation}
We evaluated Memory Reviver's technical performance using twelve photo collections provided by PVI. The evaluation focused on two aspects: (1) the scene segmentation performance and (2) the content extraction accuracy.

\textit{\textbf{Materials.}} We constructed the \emph{Memory Tree} \footnote{The Memory Tree data is provided in the supplementary materials.} from the twelve photo collections assigned to Memory Reviver in Section \ref{sec:user_evaluation_method}. These collections covered various themes, including trips and ceremonies, with each collection containing between 21 and 68 photos.

\textit{\textbf{Scene Segmentation Performance.}} To evaluate the scene segmentation, we compared our system's segmentation points with those independently marked by two researchers (Coders M and N). We used the Jaccard index to measure the similarity between each set of segmentation points. The agreement rates were 68\% between Coders M and N, 76\% between the system and M, and 65\% between the system and N. These results show that our system's agreement with human coders was similar to the agreement between human coders. When disagreements occurred, they were mainly due to differing levels of detail (e.g., a single scene by the sea vs. two distinct scenes for walking and eating by the sea).

\textit{\textbf{Content Extraction Accuracy.}} We annotated inaccuracies by reviewing all photos and statements in the \emph{Memory Tree}. A statement is considered inaccurate if it does not match the photos. One researcher labeled all the data, and a second researcher reviewed the labels to ensure reliability. We calculated the accuracy rates by dividing the number of correct statements (i.e., statements without inaccuracies) by the total number of statements. The accuracy rates were 96.4\% for storylines, 95.2\% for scene activities, and 90.8\% for scene details. The major error cases were the false identification of objects (14 times), texts (9 times), and human genders (5 times). These errors occurred due to the hallucination of GPT-4V \cite{yang2023gpt4v}.
\section{User Evaluation}
We conducted a user study with 12 PVI to evaluate the effectiveness of Memory Reviver. Our evaluation focuses on three aspects: (1) reminiscence experience, (2) understanding of a photo collection, and (3) conversational experience.

\subsection{Methods}\label{sec:user_evaluation_method}
In a within-subject study, participants used Memory Reviver and a baseline chatbot to reminisce with personal photo collections.

\subsubsection*{\textbf{Participants}} 
We recruited 12 PVI (six males, six females) who hoped to reminisce with photo collections (P7-P18, Table~\ref{tab:demographics}). These participants were recruited from an online support community. Their ages ranged from 26 to 52 (mean = 33.9, SD = 7.4). All participants were either legally blind or totally blind, and they utilized image accessibility tools such as \textit{Be My AI} \cite{be_my_eyes} and screen readers. Participants had experiences with chatbots such as ChatGPT, Gemini, and Siri. P7 and P8 took part in the formative study.

\subsubsection*{\textbf{Photo Collections}} 
All participants agreed to share personal photos for study purposes. To ensure study control, each participant provided two collections meeting the following criteria: (1) the two collections documented two distinct events; (2) both events shared the same theme (e.g., family trips) and were equally significant to the participant; (3) the difference in the number of photos between the two collections was less than five; and (4) participants had limited memory of the photos and the events.

To ensure generalizability, the collections provided by different participants covered a wide range of cases: (1) their themes included trips, gatherings, ceremonies, and conferences, and (2) the number of photos in each collection ranged from 21 to 68 (mean = 37.2).

\subsubsection*{\textbf{Baseline}}
To assess the effect of integrating user insights into Memory Reviver, we used the naïve chatbot (used in the formative study, Figure~\ref{fig:initial}) as our baseline.

\subsubsection*{\textbf{Task and Procedure}} We first gathered the participants' demographics and asked about their current practices of reminiscing with photos. Following this, the participants received a 10-minute tutorial on both Memory Reviver and the baseline chatbot using a fixed set of public images. They conducted free-form conversations with each chatbot. After the tutorial, the participants took part in two reminiscence sessions. 

The \textbf{task} in each session was to use a chatbot to reminisce with a photo collection, with the two goals identified in the formative study (section~\ref{sec:formative_goals}): (1) exploring all the scenes and (2) recalling associated memories. The order of the chatbots was counterbalanced and the photo collections were randomly assigned to participants. Each session was conducted as follows: (1) First, participants verbally composed a pre-trial \textbf{memory narrative}, prompted by the folder name of the photo collection. They were instructed to recall all memories associated with the photos. To ensure study control, no interaction with or feedback from the experimenter was provided during this narrative. (2) Next, the participants engaged in free-form conversations with the assigned chatbot. The conversation continued until the participants had no more replies. (3) After the conversation, participants composed a post-trial memory narrative using the same instructions as the pre-trial narrative. They then received a post-trial survey, which included the ratings in Figure~\ref{fig:ratings}. All ratings were on a 7-point Likert scale.

There was a 5-minute break between each session. At the end of the study, a semi-structured interview was conducted to collect participants' feedback on three aspects: (1) reminiscence experience; (2) photo collection understanding; and (3) conversational experience. The study lasted 2.5 hours and was conducted via Zoom in a one-on-one session. Both chatbots operated on the experimenter's desktop and received participants' verbal input directly via Zoom. The participants were compensated 200 CNY for their time.

\subsubsection*{\textbf{Metrics}} Informed by the two goals identified in the formative study (section~\ref{sec:formative_goals}), We adopted two task metrics:

(1) \textbf{Memory Ratio} evaluates the effectiveness of the chatbot in helping users recall memories. Specifically, \textit{Memory Ratio = (Word count of post-trial narrative) / (Word count of pre-trial narrative)}. The use of word count to measure memory recall was commonly adopted in prior studies \cite{axtell_design_2022,peesapati_pensieve_2010}.

(2) \textbf{Scene Coverage} evaluates whether the chatbot fulfilled the users' goal of exploring all the scenes in a photo collection. Specifically, \textit{Scene Coverage = (Number of scenes discussed) / (Number of all scenes)}. All photo collections were segmented into scenes using the method in Figure ~\ref{fig:4_strategy}. For Memory Reviver, a scene was considered discussed if the scene was selected for reply generation. For the baseline chatbot, a scene was considered discussed if at least one photo within the scene was selected for reply generation.

\begin{figure}[!hbtp]
    \centering
    \includegraphics[width=0.99\linewidth]{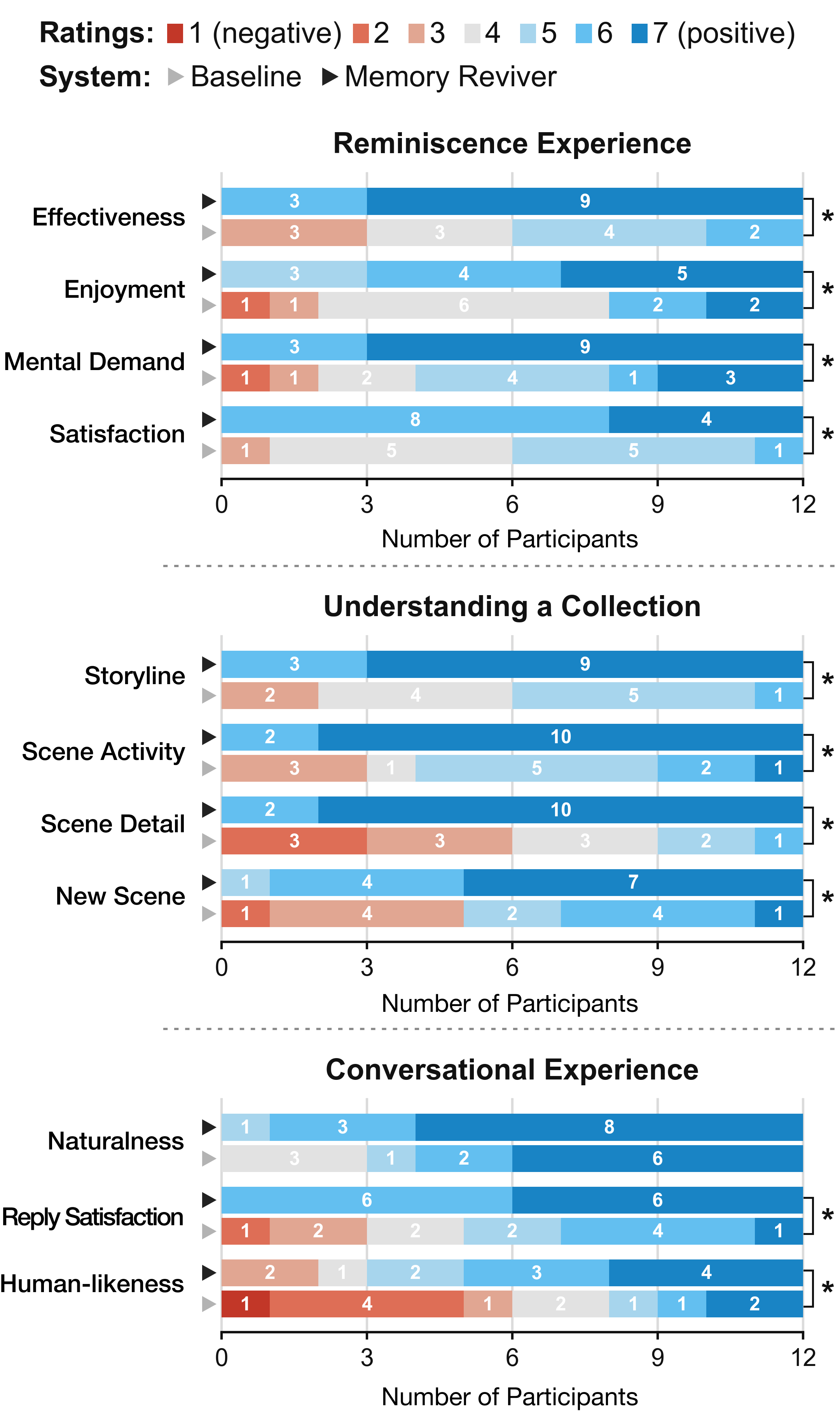}
    \caption{Distributions of the ratings for Memory Reviver and the baseline (1=negative, 7=positive). The asterisks indicate the statistical significance as a result of the Wilcoxon signed-rank test (* denotes p < 0.01).}
    \label{fig:ratings}
\end{figure}

Besides task metrics, we used subjective ratings to measure user experience in three aspects, as shown in Figure~\ref{fig:ratings}.

\subsubsection*{\textbf{Analysis}}
We recorded the study audio, participants' memory narratives, their chat history with the chatbots, and their subjective ratings. The memory narratives were transcribed verbatim as they were spoken to tally their word count. The interview audio was transcribed and categorized according to the three aspects of the interviews. Based on the data, we report our findings.

\subsection{Results}
We report results in three aspects: (1) reminiscence experience, (2) understanding a photo collection, and (3) conversational experience.

\subsubsection{\textbf{Reminiscence Experience}}
Memory Reviver was shown to be effective in facilitating memory recall and supporting enjoyable reminiscence. Moreover, we found that Memory Reviver even elicited strong positive emotions and in-depth self-reflection from some participants. The detailed results are as follows:

\begin{figure}[!htbp]
    \centering
    \includegraphics[width=1.0\linewidth]{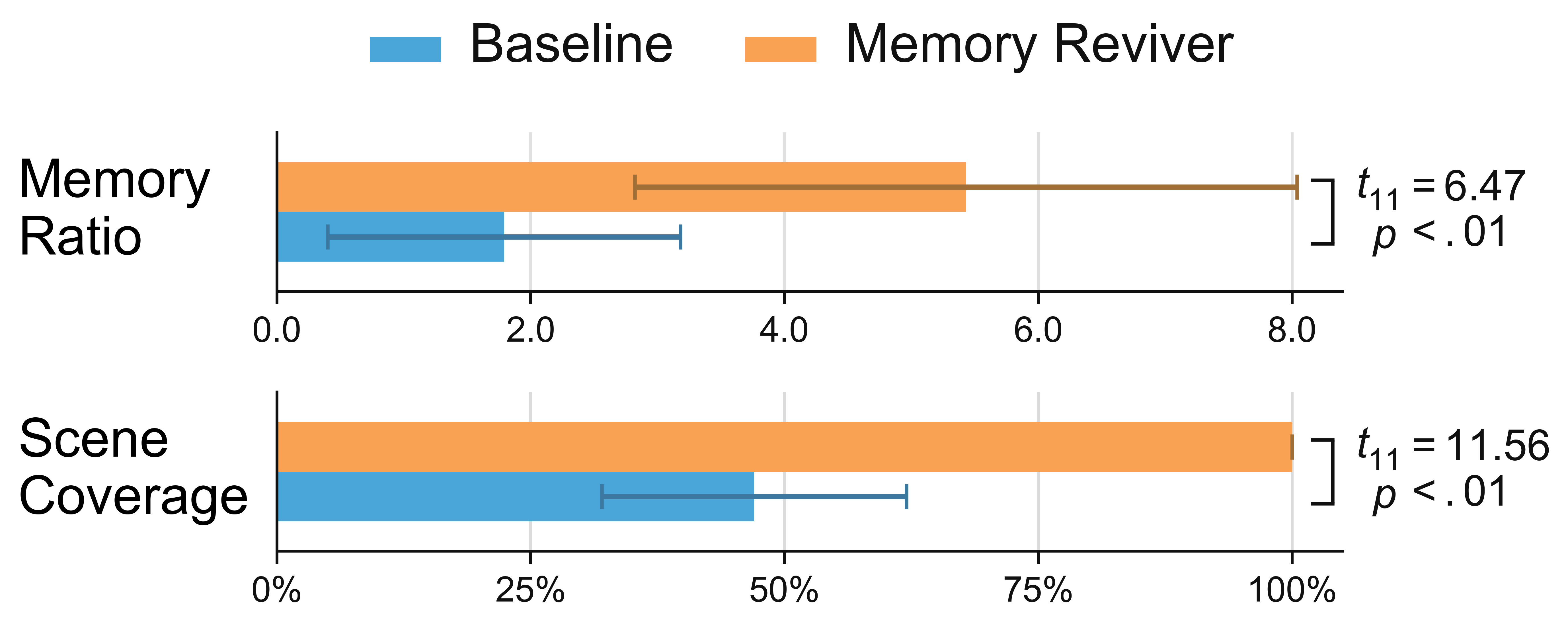}
    \caption{Task Performances in the evaluation study. Paired t-test was used for significance analysis.}
    \label{fig:metrics}
\end{figure}

\textbf{First, Memory Reviver was more effective in aiding in memory recall than the baseline}. 
Memory Reviver achieved a memory ratio of 5.43, indicating that participants narrated an average of 5.43 times more words in their post-trial memory narratives than in their pre-trial narratives. This result significantly surpassed the baseline's memory ratio of 1.79 ($\mu=5.43, \sigma=2.61$ vs. $\mu=1.79, \sigma=1.39$; $t_{11}=6.47, p<.01$). Participants rated Memory Reviver as more effective in helping them fully recall past memories ($\mu=6.75, \sigma=0.45$ vs. $ \mu=4.42, \sigma=1.08$; $Z=-3.08, p<.01$). They attributed their effective memory recall not only to the well-organized information presented by Memory Reviver but also to the low-mental-demand experience of being guided to relive past moments. As P12 noted, ``\textit{The guidance was especially useful in evoking long-lost memories... I felt fully immersed in my past.}''.
% participants narrated an average of 5.43 times more words in their post-trial memory narratives than in their pre-trial memory narratives, which is significantly higher compared to the memory ratio of 1.79 for the baseline

\textbf{Second, Memory Reviver delivered a more enjoyable experience than the baseline}. Participants rated Memory Reviver higher in enjoyment ($\mu=6.17, \sigma=0.83$ vs. $ \mu=4.58, \sigma=1.56$; $Z=-2.70, p<.01$), lower in mental demand ($\mu=6.75, \sigma=0.45$ vs. $ \mu=5.00, \sigma=1.60$; $Z=-2.68, p<.01$), and with higher overall satisfaction ($\mu=6.33, \sigma=0.49$ vs. $ \mu=4.50, \sigma=0.80$; $Z=-3.17, p<.01$). During the exit interview, all participants expressed a preference for using Memory Reviver over the baseline for future reminiscence, because it is more engaging (mentioned 12 times) and less mentally demanding (9 times). Interestingly, Participants likened their experiences with Memory Reviver to entertaining activities such as ``\textit{watching a movie}'' (P9, P17), ``\textit{taking a tour}'' (P15), or ``\textit{playing a side-scrolling game}'' (P11). In contrast, they associated the baseline with tasks such as ``\textit{retrieving information}'' (P13), ``\textit{taking a test}'' (P12), or ``\textit{finding ways in a maze}'' (P11). Participants ascribed this enjoyable and low-mental-effort experience to both the easy understanding of photo collections (mentioned 12 times) and the natural conversational experiences (9 times).

\textbf{Surprisingly, Memory Reviver elicited strong positive emotions and in-depth self-reflection for several participants}. Five participants reported experiencing strong positive emotions after reminiscing with Memory Reviver. For example, P10 stated, ``\textit{I've never imagined my memories can be revived so vividly. It made my eyes well up with tears of joy.}'' Seven participants reported reflecting on life experiences beyond photos, such as career planning and family relationships. Participants attributed their in-depth reminiscence with Memory Reviver to the low mental demand required to understand the photos, allowing them to ``\textit{fully engage in reminiscence}'' (P11). The implications of these findings are further discussed in section~\ref{sec:in-depth}.

\subsubsection{\textbf{Understanding a Photo Collection}}\label{sec:evaluation_photo_collection}
We examined whether Memory Reviver could support the users' goal of exploring all the scenes in a photo collection. We also evaluated the effectiveness of Memory Reviver in supporting the four information needs identified in the formative study. The results are as follows:

\textbf{All participants explored 100\% of scenes when using Memory Reviver}, which is significantly higher than the 47\% scene coverage achieved with the baseline ($\mu=100\%, \sigma=0\%$ vs. $\mu=47\%, \sigma=15\%$; $t_{11}=11.56, p<.01$). This indicates that Memory Reviver successfully enabled participants to explore all scenes in a collection. Participants ascribed this 100\% scene coverage of Memory Reviver to the proactive guidance: ``\textit{It listed all the scenes in the beginning, and actively guided me to explore each one. With such guidance, I felt like I was taking a tour and couldn't wait to see them all.}'' (P15). In contrast, when using the baseline, participants all ended the chat after several unsuccessful attempts to discover new scenes, as ``\textit{It often repeated what we've already discussed.}'' (P12).

\textbf{Participants rated Memory Reviver as significantly higher in fulfilling their information needs}, including a clear storyline ($\mu=6.75, \sigma=0.45$ vs. $ \mu=4.42, \sigma=0.90$; $Z=-3.11, p<.01$), easy recall of past activities ($\mu=6.83, \sigma=0.39$ vs. $ \mu=4.75, \sigma=1.29$; $Z=-2.88, p<.01$), easy exploration of scene details ($\mu=6.93, \sigma=0.39$ vs. $ \mu=3.58, \sigma=1.31$; $Z=-3.07, p<.01$), and easy discovery of new scenes ($\mu=6.50, \sigma=0.67$ vs. $ \mu=4.58, \sigma=1.68$; $Z=-2.97, p<.01$). 
Their feedback is as follows:

\textbf{1. A Clear Storyline}: All participants praised the storyline for helping them recall past memories. For example, P9 noted ``\textit{The storyline was so clear that it felt like watching a movie of my past.}''. Participants also highlighted the importance of organizing the scenes in chronological order: ``\textit{If it had ignored the timeline and grouped the photos by people or objects, I couldn't have recalled my past experiences so clearly.}'' (P14). Conversely, a lack of a well-organized storyline in the baseline resulted in a higher mental demand: ``\textit{It's like taking a test. I have to manually piece together all those memory fragments in my mind, which is quite hard.}'' (P11).

\textbf{2. Scene Activities that Aided in Memory Recall}: Participants noted that the scene activities (i.e., ``\textit{who, what, when, and where}'') effectively helped them recall past activities: ``\textit{Instead of just describing the contents and requiring me to guess what I did, it directly predicted my past activities. These descriptions instantly brought me back to the moment.}'' (P7). Participants particularly praised Memory Reviver's ability to provide supporting reasons: ``\textit{It told me the place might be either a cafe or a canteen according to the coffee cup on the table. Not even a friend has ever given such precise predictions.}'' (P13). Some participants even stated that they learned new strategies to understand their photos: ``\textit{I was pretty surprised when it told me `the name of the place might be ... according to the text on the door.' It taught me new ways to find clues in my photos.}'' (P9). In contrast, when using the baseline, participants typically asked several consecutive questions to recall their past activities, and as a result, they noted, ``\textit{It was time-consuming because I needed to figure out how the photos related to my past memories all by myself.}'' (P15).

\textbf{3. Scene Details that Enriched Memories}: Participants expressed that the active presentation of scene details gave them delightful surprises and enriched their memories. As P15 noted, ``\textit{I had the impression of taking photos there, but I had no idea so many beautiful buildings and plants were captured behind me. It truly enriched my memories.}''. The active presentation of new details made some participants feel ``\textit{warm}'' (P10, P15). Besides enriching their memories, they also mentioned that the progressive presentation of details reduced their cognitive load: ``\textit{While some tools may offer highly detailed descriptions, I often find them overwhelming. This chatbot (Memory Reviver) solved this issue for me.}'' (P13). Conversely, with the baseline, participants felt like ``\textit{retrieving information}''. As P8 noted, ``\textit{It only answered my questions, but how could I continue asking questions if I'm not learning anything new?}''.

\textbf{4. New Scene Suggestions that Reduced Mental Efforts}: Participants stated that the scene suggestions reduced their mental efforts: ``\textit{I didn't need to keep track of the discussion progress myself. I simply let it take the lead so that I could fully immerse myself in my memories.}'' (P18). Additionally, participants also noted that the summary of the current scene before suggesting the next scene was especially helpful (e.g., Memory Reviver: ``\textit{We have talked about all the key contents in the current scene: ... Do you want to proceed to the next scene?}''), because ``\textit{such a summary made me feel confident that I didn't miss out anything interesting.}'' (P7).

\subsubsection{\textbf{Conversational Experience}}\label{sec:evaluation_conversation} 

Participants generated 338 inputs when using Memory Reviver and 167 inputs with the baseline. The average response time of the chatbots was 14.6 seconds for Memory Reviver (SD=4.8, MAX=27.9) and 13.8 seconds for the baseline (SD=4.5, MAX=31.6). Regarding the conversation flow, eight participants explored all the scenes from the first to the last, and the other four participants actively switched to a specific scene (P7: 2 times; P12, P14, P17: 1 time) because ``\textit{a new scene just came into my mind}'' (P7, P14, P17) or ``\textit{I hoped to ask more about a previous scene}'' (P7, P12). We assessed the conversational experience in three aspects: (1) whether the proactive guidance of Memory Reviver would lead to unnatural conversation flow, a common concern in proactive chatbot design \cite{tallyn2018ethnobot,chabot_interact}; (2) users' satisfaction with the chatbot's replies; and (3) the chatbot's ability to simulate a human-like experience. Based on users' ratings and feedback, we have the following findings:

\textbf{First, Memory Reviver delivers a natural conversation flow by balancing proactive guidance with user freedom.} Participants rated Memory Reviver higher on average than the baseline in terms of providing a natural conversation flow ($\mu=6.58, \sigma=0.67$ vs. $ \mu=5.92, \sigma=1.31$; $Z=-1.63, p=0.10$), although no statistically significant difference was identified. They attributed this natural flow to the sense of control they had over the conversation with Memory Reviver: ``\textit{All of its guidance made me more aware of the photos. Even if some suggestions didn't interest me, I could simply ask about what I wanted to know. I was in control of the conversation.}'' (P16). Similarly, P7 noted, ``\textit{It had a default storyline, but I had total freedom to switch to another scene.}''. In contrast, participants expressed lacking control when using the baseline. As P7 noted, ``\textit{Although it seemed that we can freely communicate with the chatbot (the baseline), I often felt lost because I didn't know where our conversation was headed.}''. Collectively, findings suggest that Memory Reviver delivered a natural conversation flow by achieving a balance between providing proactive guidance and allowing users the freedom to lead the conversation.

\textbf{Second, Memory Reviver provided satisfactory replies by actively introducing new information.} Memory Reviver significantly outperformed the baseline regarding reply satisfaction ($\mu=6.50, \sigma=0.52$ vs. $ \mu=4.75, \sigma=1.54$; $Z=-2.54, p<.01$). Participants consistently attributed Memory Reviver's higher reply satisfaction to its active presentation of new information. For example, P10 noted, ``\textit{It (Memory Reviver) not only responded to my words but also mentioned something new in the photos. This kept our conversation going.}'' In contrast, participants found the baseline to provide limited information after several rounds of discussion: ``\textit{It often got stuck with a similar topic repeatedly. It never said `Oh, I had something new for you.'. This made me feel bored.}'' This finding highlights the importance of introducing new information to deliver a satisfying conversational experience.

\textbf{Third, Memory Reviver delivered a human-like experience.} Participants rated Memory Reviver as significantly more human-like than the baseline ($\mu=5.50, \sigma=1.51$ vs. $ \mu=3.75, \sigma=2.09$; $Z=-2.70, p<.01$) mainly due to its proactive guidance. As P15 noted, ``\textit{It's like a friend who is willing to guide me through the photos... It also learned from my past experiences. I felt warm.}''. During their chat with Memory Reviver, nine participants spontaneously shared their past experiences with Memory Reviver, because ``\textit{It elicited my hope to share my past experiences.}'' (P18). For example, P11 shared his anecdotes with Memory Reviver: ``\textit{You know what? We got lost during that trip ...}''. Additionally, eleven participants (all except P17) highlighted that their experiences with Memory Reviver surpassed any previous reminiscence with sighted friends. They credited this to the sufficient details provided by the chatbot (P10: ``\textit{No sighted friends have the patience to provide so many details.}'') and its dedicated services (P15: ``\textit{Chatbots always prioritize our needs, but friends may not.}''). However, participants noted that Memory Reviver's tone and response time (mean = 14.6 seconds) still presented barriers to achieving a human-like experience.
% The one exception, P17, had a partner who were experienced in writing audio descriptions.

\subsubsection{\textbf{Concerns and Improvements}}\label{sec:concerns_improvements}
We report the concerns and suggested improvements identified during the evaluation study in the following three aspects:

\textbf{(1) Incorrect Information}: Both chatbots provided incorrect information (Memory Reviver: 19 out of 338 replies; the baseline: 12 out of 167 replies), including false identification of users, visual details, and hallucinations. These errors were counted in real time by the experimenter and confirmed after the trial. Most of the errors were unnoticed during the conversation because ``\textit{I had little memory of the photos, so I accepted everything it said.}'' (P15). A few errors were noticed due to mismatches with their memories (P14: ``\textit{I never had a wrist-watch!}'') or inconsistencies across consecutive rounds of reply (P14: ``\textit{It gave me two different texts.}''). Moreover, we found that \textbf{slightly different descriptions of colors and shapes could lead to confusion}. For example, P11 noted ``\textit{The chatbot mentioned me wearing a pink dress, then a light pink one. Did I change clothes?}''). This highlights the importance of using consistent descriptions for identical objects.

After the trial, the experimenter corrected any misinformation for the participants. Eleven participants expressed tolerance of occasional errors because ``\textit{The joy of reminiscence outweighed the errors.}'' (P13). One participant (P17) cautioned against any errors because ``\textit{I only keep photos of memorable events. If the chatbot cannot be accurate, I'd prefer to reminisce with my partner.}''.

\textbf{(2) Subjective Information:}
All participants showed high acceptance of using AI-generated storyline and activity descriptions for reminiscence. While such information tends to be subjective and may not perfectly align with users' past experiences, participants expressed few concerns. They highlighted two main reasons: First, they believed that no other individuals could perfectly describe their past experiences: ``\textit{Even different friends describe my photo collections differently.}'' (P11). Second, they mentioned that the supporting reasons effectively alleviated their concerns about subjectivity (e.g., ``\textit{You seemed to be in a zoo because of the `no feeding animals' sign.}''). This finding indicates that combining factual contents with subjective descriptions is a potential solution to reduce PVI's concerns about subjectivity, a common issue mentioned in prior works \cite{stangl_going_2021,stangl_person_2020, jung_so_2022}.

\textbf{(3) Future Improvements of Memory Reviver:}
Participants provided suggestions for improving Memory Reviver. The most common suggestion (mentioned 7 times) was for Memory Reviver to have long-term memories across different reminiscence sessions. For example, P13 noted, ``\textit{I want it to learn that this dog is my guide dog not only for this conversation but for all future ones.}''. Five participants noted that Memory Reviver sometimes suggested new information that had already been mentioned earlier in the conversation. To address this issue, future improvements should focus on filtering out redundant information that has already been discussed before adding the proactive guidance. Four participants suggested that Memory Reviver should be able to respond to questions related to multiple scenes (e.g., ``\textit{Is the dog on the beach the same as the one on the grass?}''). Additionally, three participants suggested customizing the role of the chatbot or its narrative styles, which we discuss in section~\ref{sec:future_improvements}.

\section{Discussion}
We reflect on findings from the design and evaluation of Memory Reviver and discuss the implications for future design.

\subsection{Personalization of Memory Reviver}\label{sec:future_improvements}
Based on the evaluation study, we identify several future directions to tailor Memory Reviver to different users' preferences.

\textbf{Personalize responses using interaction histories.} While PVI could ask diverse visual questions \cite{stangl_going_2021,jung_so_2022}, we noticed that the same participant in the user study tended to ask similar questions when viewing photos in a collection. For instance, P16 frequently inquired about facial expressions (14 out of 21 questions), while P13 often asked about body postures (8 out of 17). Our preliminary observation suggests that this tendency might be related to both the photo themes (e.g., a family trip) and their personal preferences. This indicates an opportunity to learn from user preferences and adjust subsequent responses using online learning methods \cite{hoi2021online}.

\textbf{Prioritize information based on user contexts.} In our user study, participants indicated that their reminiscence interests might depend on their current contexts, such as locations (P11: ``\textit{When revisiting a country, I like to review photos taken there previously.}''). This suggests that Memory Reviver could prioritize content based on user contexts like location \cite{mcgookin_reveal_2019} and time \cite{peesapati_pensieve_2010}.

\textbf{Reduce the frequency of suggestions for older adults.}
Currently, Memory Reviver provides proactive guidance in every reply. While eleven participants found this frequency appropriate, P9 (aged 52) occasionally experienced an interruption of thought due to the new information in the suggestions. This could be linked to age-related cognitive declines \cite{czaja2006factors}. Given that older adults often have a strong desire for reminiscence \cite{peesapati_pensieve_2010}, it's essential to customize the frequency of suggestions to their needs.

\textbf{Tailor the role of the chatbot.}
Participants in our evaluation study had differing views on the role of the chatbot. Four participants envisioned the chatbot as a friend with a consistent personality: ``\textit{I hope it could chat with me about its own experiences.}'' (P10). Conversely, eight participants preferred the chatbot to act as an information provider, as they viewed reminiscence as a personal experience: ``\textit{It serves to provide information and I'll reminisce by myself.}'' (P17). This indicates the role and personality of the chatbot \cite{park2023generative,wang2024survey} could be further customized.

\textbf{Customize description styles.}
Participants held diverse preferences regarding the description styles of the chatbot. For example, P15 expected the chatbot to provide subjective descriptions such as ``\textit{You seemed really happy.}'', while P17 preferred more objective descriptions. This indicates that the description styles could be further customized by leveraging insights from prior works \cite{morris_rich_2018}.

\subsection{Privacy and Ethical Considerations}

Our research involved using personal photos to elicit personal memories. To address the associated privacy concerns, we carefully took measures in both user studies. First, we informed participants of the privacy risks associated with submitting photos to a third-party service and obtained their consent. Second, we removed all metadata from the photos and ensured that data used in API calls would not be used for training purposes. Third, we submitted data deletion requests after the studies. We acknowledge that some privacy risks, such as data leaks by third-party services, remain.

To eliminate the privacy risks, future work should focus on using local visual language models on mobile devices \cite{chu2024mobilevlm,apple_ai} to enhance data privacy. Moreover, reminiscence may trigger memories of negative experiences \cite{isaacs2013echoes}. To minimize such negative effects, future systems should enable users to filter out certain content before starting reminiscence sessions.

\subsection{Design Implications}

\subsubsection*{\textbf{Photo Reminiscence Support}}\label{sec:in-depth}

During the user evaluation, participants mentioned the unique benefits of reminiscing with a chatbot: offering patient and dedicated services, reducing social concerns, and promoting self-reflection. Notably, seven participants spontaneously engaged in self-reflection after using Memory Reviver. For example, P7 contemplated career planning, P8 decided to reconnect with long-lost friends, P13 resolved to spend more time with her parents, and P16 expressed the intent to resume playing the piano. Their self-reflection indicated the depth of reminiscence \cite{isaacs2013echoes}. Participants noted that such depth was rare in their previous reminiscence with photos. We identified three contributing factors to this in-depth experience. (1) The \textbf{ease of understanding a photo collection} allowed them to engage in self-reflection. As P7 noted, ``\textit{I naturally reflected on my life choices because understanding many photos was no longer tedious.}''. (2) The \textbf{rich details} helped them deeply connect with and reflect on past experiences: ``\textit{After hearing how splendid the music hall was, I wanted to perform there again.}'' (P16). (3) The \textbf{independent reminiscence} facilitated their self-reflection. As P8 noted, ``\textit{When I reminisce with friends, it's more about chit-chat than reflection.}''.

Collectively, our study reveals the significant potential of photos in facilitating self-reflection for PVI. This finding suggests that future photo reminiscence support should not only focus on helping PVI understand photo contents \cite{zhao2017effect,yoo_understanding_2021,jung_so_2022}, but also aim to engage them in self-reflection \cite{bentvelzen2021technology}. By alleviating the cognitive load of understanding photos, providing rich details, and supporting independent reminiscence, accessibility tools can help PVI deeply engage with photos and derive psychological benefits.

\subsubsection*{\textbf{Chatbot Design for PVI}}
Our study revealed that PVI aimed to fully explore a photo collection during their conversations with a chatbot. This objective differs from chat-based information retrieval \cite{levy2024chatting} or question answering \cite{vqa}, as it requires users to actively organize and synthesize large amounts of information. However, achieving this task poses challenges due to the limited memory storage capacity of auditory channels \cite{card2018psychology}. To address this challenge, our approach is to (1) distill PVI's information needs by observing their natural conversations with a chatbot, (2) organize the essential information into a clear structure, and (3) gradually present information to users \cite{morris_rich_2018,peng2023slide,huh2022cocomix}. User evaluation demonstrated the effectiveness of our design.

Similar to this task, PVI also need to comprehend large amounts of information for tasks such as reading \cite{li2022enhancing} and web browsing \cite{pucci2023defining}. Drawing upon our findings and established guidelines \cite{chabot_interact, nielsen2005ten}, we derive implications for designing accessible chatbots aimed at supporting the comprehension of large amounts of information as follows: (1) Identify users' information needs and organize information into a clear structure, (2) Offer an overview to help users quickly skim the information, (3) Present information progressively to alleviate cognitive load, 
% (4) Keep users aware of their progress (e.g., ``\textit{This is the fifth scene.}''), 
and (4) Summarize information regularly to mitigate users' fear of missing out (e.g., Memory Reviver provides a summary of the current scene before suggesting a new scene.). These implications extend prior literature on accessible chatbots for PVI \cite{choi2020nobody, pucci2023defining, branham_reading_2019, pradhan_accessibility_2018, abdolrahmani_siri_2018}, offering practical insights for designing chatbots tailored to the needs of PVI when consuming large amounts of information.

Additionally, when designing assistive chatbots for real-time scenarios such as outdoor navigation \cite{xu2020virtual,yang2021lightguide,liu2021tactile} and shopping \cite{boldu2020aisee}, it is vital to avoid overwhelming users with excessive information \cite{boldu2020aisee}. Our strategy of presenting information in an ``overview first, details on demand'' manner could help reduce the associated cognitive load in these scenarios.

\subsection{Limitations and Future Work}
We summarize the limitations and future work in two aspects: (1) the scope of Memory Reviver, and (2) limitations of user studies.
\subsubsection*{\textbf{Scope of Memory Reviver}}\label{discussion_scope}
Memory Reviver is designed to help PVI reminisce with a photo collection using a chatbot. Its scope is bounded by the use of conversational user interfaces, the types of photo collections it can process, the reminiscence needs it caters to, and the performance of its underlying models. 

First, Memory Reviver leverages natural conversation with a chatbot to facilitate reminiscence. Future work should explore additional design possibilities, such as utilizing multi-modal input (e.g., touch-based image exploration \cite{huh2023genassist, nair_imageassist_2023}) and multi-sensory feedback (e.g., auditory and tactile feedback \cite{yoo_understanding_2021,harada2013accessible, wang2024virtuwander}), to further enhance the reminiscence experience.

Second, Memory Reviver is tailored for photo collections associated with specific events like trips, family gatherings, ceremonies, and stage performances (see Table~\ref{tab:demographics}). Such collections typically comprise dozens to hundreds of photos that can be segmented into multiple scenes. To extend Memory Reviver’s ability in processing larger and more diverse photo libraries \cite{chen_exploring_2023}, future work should explore methods to automatically select photos for reminiscence \cite{apple_ai,mcgookin_reveal_2019} and use metadata to filter out outliers like screenshots. Additionally, we noticed that users' inquiries may be influenced by the themes of photo collections. Therefore, fine-tuning proactive suggestions based on the specific themes of photo collections could be a potential direction for future work.

Third, Memory Reviver is designed to accommodate the reminiscence needs of fully exploring a photo collection. Currently, it restricts exploration on a scene basis and does not support cross-scene queries, which we view as promising future work. Moreover, users may have other needs when interacting with a photo collection, such as retrieving a specific photo \cite{harada2013accessible,jung_so_2022}. While the specific solution may not be directly applicable, our methodology of delivering proactive guidance according to user needs can be leveraged.

Fourth, Memory Reviver inherits the limitations of its underlying models. The use of GPT-4V \cite{yang2023gpt4v} occasionally introduced image recognition errors and hallucinations, which could be mitigated by incorporating specialized models (e.g., precise text recognition \cite{biten2022latr}) and model advancements (e.g., alleviating hallucinations \cite{biten2022let}).

Fifth, Memory Reviver currently employs rule-based methods to select proactive suggestions from pre-extracted information (i.e., the \emph{Memory Tree}). Future works include improving the suggestion methods and customizing pre-extracted information according to diverse user preferences.

\subsubsection*{\textbf{Limitations of User Studies}}

First, participants in our evaluation study reminisced with their personal photo collections. While we attempted to control for variations between collections, differences could not be completely eliminated. Second, the majority of participants in our user studies were young adults, with only one older participant (P9, aged 52). As such, the effectiveness of Memory Reviver for older adults requires further investigation. Third, we did not evaluate the long-term effects of Memory Reviver in supporting daily reminiscence, which we see as potential future work.

\section{Conclusion}

Memory Reviver is a proactive chatbot designed to assist people with visual impairment in reminiscing with a photo collection. It addresses two key challenges hindering
effective reminiscence with a chatbot: the scattering of information and a lack of proactive guidance. 
Through user evaluation, we demonstrated the effectiveness of Memory Reviver in facilitating engaging reminiscence, enhancing understanding of photo collections, and delivering natural conversations. We identified future directions to personalize Memory Reviver according to diverse user needs and distilled implications for photo reminiscence support and chatbot design for PVI. We hope this work will offer useful insights into the design of accessible chatbots and inspire researchers to develop tools that facilitate enjoyable and in-depth reminiscence.

\begin{acks}
The authors would like to thank the participants for their support during the studies. Additionally, we thank the reviewers for their constructive feedback. We thank Prof. Chun Yu from Tsinghua University and Prof. Guanhong Liu from Tongji University for their methodological guidance. We thank Qiyue Cai from Arizona State University for introducing the structure of autobiographical memory, which inspired the design of \emph{Memory Tree}. This work is partially supported by the Research Grants Council of the Hong Kong Special Administrative Region under General Research Fund (GRF) with Grant No. 16214623.
\end{acks}

\balance
\bibliographystyle{ACM-Reference-Format}
\bibliography{_references}

\appendix
\newpage\onecolumn
\section{Tables}
Table~\ref{tab:demographics} shows the demographics of the participants. Table~\ref{tab:subjective_ratings} shows the subjective ratings in the evaluation study.

\begin{table*}[!htbp]
  \centering
  \caption{Demographic information of the participants.}
  \renewcommand\arraystretch{1.2}
  \setlength{\tabcolsep}{1mm}{
      \begin{tabular}{c c c c c l l l}
        \hline
        \textbf{PID} & \textbf{Age} & \textbf{Gender} & \textbf{Visual Condition} & \textbf{Onset} & \textbf{Image Accessibility Tools} & \textbf{Chatbot Usage} & \textbf{Themes of Photos in Study} \\
        \hline
        P1 & 29 & M & Legally blind & Acquired & Seeing AI, Talkback & Huawei Xiaoyi & Family gatherings \\
        P2 & 45 & F & Totally blind & Congenital & VoiceOver & Siri & Family trips \\
        P3 & 26 & M & Totally blind & Acquired & Be My AI, VoiceOver & ChatGPT, Siri & Family trips \\
        P4 & 27 & F & Legally blind & Congenital & Be My AI, VoiceOver & Siri & Stage performances \\
        P5 & 38 & M & Totally blind & Congenital & Be My AI, VoiceOver & Siri & Family gatherings \\
        P6 & 35 & F & Totally blind & Congenital & Talkback & Xiaomi xiao'ai & Ceremonies \\
        P7 & 30 & F & Totally blind & Congenital & Be My AI, VoiceOver & ChatGPT, Siri & Business trips \\
        P8 & 33 & M & Legally blind & Congenital & Be My AI, Talkback & Xiaomi xiao'ai & Trips with friends \\
        P9 & 52 & F & Totally blind & Acquired & VoiceOver & Siri & Family gatherings \\
        P10 & 30 & F & Totally blind & Congenital & Be My AI, VoiceOver & Siri & Trips abroad \\
        P11 & 32 & M & Legally blind & Acquired & Be My AI, Talkback & ChatGPT & Trips abroad \\
        P12 & 31 & M & Totally blind & Congenital & Be My AI, VoiceOver & Gemini & Conference speech \\
        P13 & 44 & F & Legally blind & Acquired & Be My AI, VoiceOver & Siri & Family trips \\
        P14 & 36 & M & Legally blind & Congenital & VoiceOver & ChatGPT, Siri & Family trips \\
        P15 & 27 & F & Legally blind & Congenital & VoiceOver & Siri & Trips with friends \\
        P16 & 26 & M & Legally blind & Congenital & Be My AI, VoiceOver & Siri & Stage performances \\
        P17 & 30 & F & Totally blind & Acquired & Be My AI, VoiceOver & ChatGPT, Siri & Family trips \\
        P18 & 36 & M & Totally blind & Congenital & Be My AI, VoiceOver & Siri & Ceremonies \\
        \hline
      \end{tabular}
    }
  \label{tab:demographics}
\end{table*}

\begin{table*}[!hbtp]
\caption{{Subjective ratings. (1=Strongly Disagree, 7=Strongly Agree. {Wilcoxon signed-rank test was used for significance analysis.})}}
~\label{tab:subjective_ratings}
\centering
\hspace{-4.5mm}
\resizebox{0.99\columnwidth}{!}{
% \resizebox{2.09\columnwidth}{!}{
\setlength{\tabcolsep}{1.32mm}{
\renewcommand\arraystretch{1.1}
\newcommand{\tabincell}[2]{\begin{tabular}{@{}#1@{}}#2\end{tabular}}
\newcommand{\hlineblack}{\specialrule{0.1em}{0em}{0em}}
\begin{tabular}{c | l | c c | c}
%\toprule
\hlineblack
{\textbf{Aspects}} & {\textbf{Participant statements}} & {\textbf{Memory Reviver}} & {\textbf{Baseline}} & {\textbf{Significance}}\\
%\midrule
\hline
\multirow{4}{*}{\parbox{2.0cm}{\centering Reminiscence\newline Experience}} & \textbf{Effectiveness}: I fully recalled memories about past events. & 6.75 (SD=0.45) & 4.42 (SD=1.08) & $Z=-3.08, p<.01$ \\
& \textbf{Enjoyment}: I felt happy when reminiscing with this chatbot. & 6.17 (SD=0.83) & 4.58 (SD=1.56) & $Z=-2.70, p<.01$ \\
&  \textbf{Low Mental Demand}: I didn't feel mentally demanded using this chatbot. & 6.75 (SD=0.45) & 5.00 (SD=1.60) & $Z=-2.68, p<.01$ \\
& \textbf{Overall Satisfaction}: I am satisfied with the reminiscence experience. & 6.33 (SD=0.49) & 4.50 (SD=0.80) & $Z=-3.17, p<.01$ \\
\hline
\multirow{4}{*}{\parbox{2.0cm}{\centering Understanding\newline a Collection}} & \textbf{Clear Storyline}: I clearly grasped the storyline of all the scenes. & 6.75 (SD=0.45) & 4.42 (SD=0.90) & $Z=-3.11, p<.01$ \\
& \textbf{Easy Recall of Activities}: I easily recalled past activities in each scene. & 6.83 (SD=0.39) & 4.75 (SD=1.29) & $Z=-2.88, p<.01$ \\
& \textbf{Easy Exploration of Details}: I easily explored the details in each scene. & 6.83 (SD=0.39) & 3.58 (SD=1.31) & $Z=-3.07, p<.01$ \\
& \textbf{Easy Discovery of New Scenes}: I easily found new scenes to talk about. & 6.50 (SD=0.67) & 4.58 (SD=1.68) & $Z=-2.97, p<.01$ \\
\hline
\multirow{3}{*}{\parbox{2.0cm}{\centering Conversational\newline Experience}} 
& \textbf{Natural Conversation Flow}: The conversation flowed naturally. & 6.58 (SD=0.67) & 5.92 (SD=1.31) & $Z=-1.63, p=0.10$ \\
& \textbf{Reply Satisfaction}: The chatbot provided satisfying replies. & 6.50 (SD=0.52) & 4.75 (SD=1.54) & $Z=-2.54, p<.01$ \\
& \textbf{Human-likeness}: I felt like I was talking to a real person. & 5.50 (SD=1.51) & 3.75 (SD=2.09) & $Z=-2.70, p<.01$ \\
%\bottomrule
\hlineblack
\end{tabular}
}
}
\end{table*}

\end{document}